# Reliable System Specification for Self-Checking Data-Paths

C. Bolchini, F. Salice, D. Sciuto
Politecnico di Milano, Dip. di Elettronica e Informazione
Milano, Italy
cristiana.bolchini@polimi.it. salice@elet.polimi.it, sciuto@elet.polimi.it

L. Pomante
CEFRIEL
Milano, Italy
pomante@cefriel.it

**Abstract**

*The design of reliable circuits has received a lot of attention in the past, leading to the definition of several design techniques introducing fault detection and fault tolerance properties in systems for critical applications/environments. Such design methodologies tackled the problem at different abstraction levels, from switch-level to logic, RT level, and more recently to system level. Aim of this paper is to introduce a novel system-level technique based on the redefinition of the operators functionality in the system specification. This technique provides reliability properties to the system data path, transparently with respect to the designer. Feasibility, fault coverage, performance degradation and overheads are investigated on a FIR circuit.*

## 1 Introduction

When embedded systems implement critical applications they should exploit a certain degree of reliability; in particular, Concurrent Error Detection (CED) capabilities are frequently a requirement, since in critical operational environments error propagation can have catastrophic effects.

In the past, a set of techniques [11, 13] has been proposed (for both combinational and sequential devices) in order to provide such CED property by acting directly at low abstraction level, toward the end of the hardware design flow. However, time-to-market constraints and the increasing complexity of modern devices make it impossible to manage the problem at such a low level and a different approach, able to raise the abstraction level, is necessary. As a matter of fact, standard design methodologies target devices by starting from system level specifications. At this description level, the system specification identifies the functional characteristics of the application without detailing implementation aspects. Because the insertion and the use of reliability methodologies significantly impacts on performance, timing, energy and area, it is necessary to transfer these aspects toward the upper levels of the synthesis flow, by adding the reliability constraints to the classical cost/performance parameters. Our goal is to define a co-design methodology able to integrate CED capabilities from the first steps of the design process, so to achieve a reliable system. The need of frameworks for reliable co-design has been already highlighted in the last years, leading to the development of approaches able to manage reliability at system level. However some of the studies presented in literature [19] focus on Design for Testability (DfT) techniques defined at system-level without considering CED approaches. Other works [4] consider such kind of property by means of a tagged specification used to drive a special two-level partitioning to identify the proper reliable design methodology to be used in the subsequent steps of the system design process. Another important work is the one presented in [6] where a task-graph based co-design environment (suited for specific application domains) provides error detection properties based on duplication or assertions. However, the use of such assertions is not transparent to the designer and a quantitative evaluation of the fault coverage cannot be computed. The proposal in [2] considers behavioral assertions to drive a high-level synthesis tool: such an approach is interesting but it is hardware oriented and the assertions involved are related only to area and performance issues (i.e. no fault tolerance). In [5] a software modification strategy is proposed for on-line fault detection; the authors define a set of rules for explicitly introducing redundancy in the high-level code, in some specific situations.

This paper presents a novel approach to reliable system design by exploiting the SystemC-Plus object-oriented features that can be applied in the more general framework proposed in [4]. The methodology is presented in detail with respect to a limited running case in order to describe both advantages and drawbacks of the approach. The methodology, as it is, has led to significant results, yet it is currently under development in order to better exploit the potential benefits of the adopted mechanisms, while minimizing the present limitations. The methodology considers SystemC-Plus descriptions as a starting point, thus giving the possibility of synthesizing the enriched specification either in hardware or in software, the proposed approach is independent of the final implementation and thus it can be exploited in a complete co-design flow (e.g. simulation, partitioning, synthesis, etc.) without any specific intervention, and so obtaining a straightforward way to reach a reliable hw/sw implementation.

This paper is organized as follows. Section 2 describes the key elements of the proposed approach, whereas the details of the defined self-checking class are introduced in Section 3. Section 4 discusses the quality of the technique, investigating fault coverage issues for several target architectures. The proposed co-design flow is presented in Section 5, with the results of the FIR benchmark application. The last section draws some conclusions and future work.

## 2 The proposed approach

In the following the key elements of the proposed approach are detailed: the specification language, the fault model, the design methodology, the target architecture and the coverage analysis.

**Specification language.** The proposed approach is based on the *SystemC-Plus* specification language [18], a proper extension of the synthesizable *Synopsys-subset of SystemC* [17], that allows full exploitation of object-oriented features (inheritance, polymorphism, etc.) in the specification, while keeping hardware and software synthesis capabilities.

**Fault Model.** The adopted fault model consists of the *single functional unit failure*, where any number of physical faults can cause one (and only one) functional unit to perform incorrectly. The considered faults affect the target platform, mining hardware and software execution units. The functional units considered in this work are those dedicated to perform the basic arithmetic and logical computations (e.g. ALUs, multipliers, shifters, etc.), i.e. those that can be mapped directly onto the operators (arithmetic, logic or bit-a-bit) used in the specification (e.g. =, +, -, *, etc.). The functional failure is modeled by means of a variable number of errors (e.g. stuck at, bit-flip, etc.) affecting the bits of the result provided by the faulty unit. Both permanent and transient and intermittent faults are covered by our approach, the latter increasingly like to occur in any integrated device, due to the technology trend that reduces size and increases operating frequency ([12]).

**Design methodology.** According to the adopted fault model, the idea is to provide the designer with the possibility of using a self-checking data type that autonomously performs controls on data resulting from basic computations. The verification of results' correctness is achieved by means of one or more functional redundancy mechanisms (the number and the complexity of such mechanisms will depend on the desired trade-off between fault coverage, cost and performance) as typically adopted in the software field (e.g. software checking [20, 15, 21, 14], and assertion [9, 1, 8]).

**Target architecture.** Because the specification is independent of the target platform, the proposed approach does not need any particular information about the final architecture. The resulting implementation (totally hardware, totally software with one or more processors, or a mix of them) depends on the adopted design flow and the constraints to be satisfied, as in any hw/sw co-design flow.

### 2.1 Coverage analysis

The effectiveness of the proposed self-checking technique is evaluated in terms of the achieved fault coverage and the additional costs. Both qualitative and quantitative analyses have been carried out. As an example, consider the z=x+y operation, controlled by the inverse operation w=z-x (in this case the controlling mechanism is w==y). Referring to the adopted fault model, two situations can be envisioned. Using a multi functional resource system and a proper allocation/scheduling policy it is possible to achieve a 100% fault coverage if different functional units perform the two operations. On the other hand, a software implementation on a monoprocessor system (or a limited resource hardware system) could lead to a solution where the same functional unit could perform both operations. In the latter case there could be faults causing undetected erroneous data affecting the fault coverage to some degree since one unit performs the operation and its control. For example, there could be a situation where, by means of an undesired error masking (or error compensation), an erroneous z' could lead to a w'=z'-x, with w' coincident to y so nullifying the error control. Section 5 dedicates particular attention to the worst-case fault coverage evaluation, investigating the probability of a fault causing consistent error compensation when performing two related operations.

In the following the + operator and its underlying hardware adder component will be used as running example, without loss of generality. The same considerations apply to any arithmetic/logical component working on integer data, in a straightforward manner; only the overall cost/performance results are here reported. The limitation to integers depends on *SystemC* ability to synthesize only this type.

## 3 Class template SCK<TYPE>

The operative element of the proposed approach is the SCK class template (it is the only *SystemC-Plus* exten-sion used with respect to synthesizable *SystemC*). The use of a class template in C++ fashion (e.g. as proposed in [10] where a very similar approach is used to prevent errors in the software design) allows the designer to adopt it with all the synthesizable basic data types by changing only the data declaration. A proper overloading mechanism of the basic operators applies the desired reliability-oriented mechanisms, while the *SystemC-Plus* features will guarantee the synthesizability property. For example, when using an `SCK<int>` data type, the *operator +* will verify the consistency of its result by means of an inverse operation (i.e. *operator -*), consequently updating the value of an error bit associated with the data itself. Operators are designed to propagate also the error bit value. As a result, we obtain a specification that includes the operations verifying the correctness of the provided results, while keeping a minimal impact on the way the system is specified.

Operator overloading is not an innovative technique, although it has been employed in the past for different goals and generally for dealing with software properties, rather than being used to detect hardware failures as in the present approach.

### 3.1 SCK interface

Figure 1 shows the interface of the class SCK (limited in this example for clarity to operators = and +). The error bit $E$ has been introduced and associated with the internal data $ID$, together with the methods used to manage such data (*GetError* and *GetID*) and the prototypes of the useful versions of the two operators. It is worth noting that the presence of the empty constructor is due to synthesis constraints.

```
template<class TypeSCK>
class SCK
{
  protected:
    TypeSCK ID;// Internal data
    bool E;    // Error

  public:
    SCK(){};    // Empty constructor

  // MANAGEMENT
  bool GetError() const;
  TypeSCK GetID() const;

  // OPERATORS OVERLOADING
  void operator=(const SCK<TypeSCK> & OX);
  void operator=(const TypeSCK & X);

  SCK<TypeSCK> operator+(const SCK<TypeSCK> & OX) const;
  SCK<TypeSCK> operator+(const TypeSCK & X) const;
  //Non-member function
  //SCK<TypeSCK> operator+(const TypeSCK & X,const SCK<TypeSCK> & OX)
};
```

**Figure 1. The SCK interface.**

### 3.2 SCK implementation

Figure 2 shows the self-checking implementation of the operator + used in the SCK class. The code shows the strategy used to provide error propagation and inverse operation check. In particular a `z=x+y` operation is checked by means of the `(z-y==x)` error control. Although a strategy has been fixed in the SCK class, it is straightforward to provide different implementations to obtain a different trade-off between cost and reliability.

```
template <class TypeSCK>
SCK<TypeSCK>
SCK<TypeSCK>::operator+(const SCK<TypeSCK> & OX) const
{
  SCK<TypeSCK> Out;
  Out.E=false;

  Out.ID=ID+OX.ID;
  if ((Out.ID-OX.ID!=ID)||(OX.E)||(E)) Out.E=true;
  return Out;
}
```

**Figure 2. Self-Checking** *operator +*.

For example, a different implementation of the + operator uses the `(z-y==x)&&(z-x==y)` control condition proving higher fault coverage and hardware cost.

A similar hidden control mechanism has been defined for the other arithmetic operators, taking also into consideration problems related to the precision of the inverse operation (for the division case); Table 1 reports the most promising overloading techniques we evaluated.

| Operators | Techniques | | |
|---|---|---|---|
| | Tech1 | Tech2 | Both |
| Add + | op2' = ris - op1 | op1' = ris - op2 | |
| ris = op1+op2 | op2 == op2' | op1 == op1' | |
| Fault Cov. | 97.25% | 98.81% | 99.11% |
| Sub - | op1' = ris + op2 | ris' = op2 - op1 | |
| ris = op1 - op2 | op1 == op1' | 0 == ris + ris' | |
| Fault Cov. | 96.85% | 94.01% | 99.58% |
| Mult + | ris' = (-op1)×op2 | ris' = op1×(-op2) | |
| ris = op1×op2 | 0 == ris + ris' | 0 == ris + ris' | |
| Fault Cov. | 96.22% | 96.38% | 97.43% |
| Div / | op1' = ris×op2 | op1' = -ris×op2 | - |
| | + (op1 % op2) | - (op1 % op2) | |
| ris = op1 / op2 | op1 == op1' | op1 == op1' | - |
| Fault Cov. | 94.33% | 97.16% | - |

**Table 1. Overloading techniques and fault coverage.**

It is worth noting that there is no available tool for evaluating the fault coverage of the final realization with respect to the on–line fault detection properties, yet the local fault coverage analysis (carried out with an exact fault detection analysis for the single arithmetic units) can be used as an estimation of the reliability level that will be achieved.

## 4 Fault coverage analysis

The effectiveness of our approach relies on the percentage of faults that may cause undetected errors when using a system with limited resources such that the nominal operation and its controlling operation cannot unveil a fault, as introduced in Section 2.1. An in-depth analysis based on the integer faulty class has been carried out. For example, using the FAddFSubIntC faulty type, a fault causing an error on the addition could produce also an error on the subtraction operation such that the fault is not detected. This consideration highlights the need for a detailed low-level analysis to evaluate the fault coverage guaranteed by the proposed approach with respect to several target architectures.

Let us consider, as an example, instruction c=a+b. As shown in Figure 2, the enhanced operator + will perform the addition a+b and will also perform the hidden operation d=c+(-a) followed by a comparison between d and b. In a fault free situation, with the exception of overflows (which are separately dealt with) d equals b; no error indication is set. On the other hand, when a fault occurs within the adder (and c=a+b produces c' instead of c) when performing the hidden operation d'=c'+(-a), one of the following situations may arise:

1. the operation is performed on a different functional unit (fault free for hypothesis), thus the resulting value d' is different from b: the fault is detected;
2. the operation is performed on the same faulty unit, in this case two are the possible outcomes:
   (a) d' differs from b: the fault either is not observable, hence d' is the correct result of the operation, or is observable and detected.
   (b) d' equals b: the fault is observable in such a way that an erroneous result is produced, but in such a way that the two errors mask the fault.

Situation (2b) is the critical one and it affects the fault coverage due to the presence of faults that the methodology cannot cover. It is worth noting that the problem arises only when the same resource is used to compute both the related operations: it is thus necessary to evaluate the exact fault coverage that is achievable by means of the proposed approach in such a situation (worst case analysis), as discussed in the following.

### 4.1 The testing environment

Fault coverage is evaluated at hardware level, by considering the target architectures and the faults included in the adopted fault model. A specific test architecture has been defined to analyze the faults masking effects with respect to the pairs of operations used within the overloading of the SCK class operators. More in detail, we simulate faults effects considering the pair of operations that are implemented in the enhanced operator (e.g., + and - for operator +) when they are performed on the same functional unit. The test architecture is independent of the actual implementation, and can be used with different technological choices, with a carry look-ahead implementation of an adder, as well as with a ripple carry implementation. Two environments have been setup, in VHDL and C. The g function, when needed, is the one used to produce the value for the dual operation (e.g. in the case of the + and - operations pair, the g function performs the 1's complement and the f function receives a 1 on the *carry-in* to work in 2's complement).

Table 2 shows how the coverage analysis has been carried out, reporting the experimental results obtained for a ripple-carry adder performing the + operation, where "Tech 1" refers to a single overloading of the operator using as control operation op2=ris-op1; "Tech 2" refers to the dual controlling operation op1 = ris-op2; the third approach ("Tech 1&2") includes both overloading strategies.

To obtain results independent of the actual implementation of the adder (e.g. FPGA, ASIC with standard or full-custom, etc.), functions f, g and faults have been modeled at the functional level (i.e. the faulty functional unit is the single full-adder in the chain composing the $n$-bit adder).

| Overloading $\Rightarrow$ | | Tech1 | Tech2 | Tech 1&2 |
|---|---|---|---|---|
| # bits | # fault situations | | | |
| 1 | 128 | 95.31% | 96.88% | 97.66% |
| 2 | 1024 | 96.88% | 98.44% | 98.83% |
| 3 | 6144 | 97.40% | 98.96% | 99.22% |
| 4 | 7808 | 97.66% | 99.22% | 99.41% |
| 8 | $16 \times 2^{20}$ | 98.05% | 99.61% | 99.71% |
| 16 | $6 \times 2^{30}$ | 98.18% | 99.74% | 99.80% |

**Table 2. Experimental results for different overloadings for** *operator* **+.**

The first column of Table 2 refers to the width of the operands used to perform the computation, ranging from 1 to 12 bits. The second column reports the total number of fault situations for each case, dependent on the number of input combinations and faults in the adder, given by the following equation:

No. of faulty situations = num_faults_1bit $\times n \times 2^{2n}$

where $n$ is the operand size, $num\_faults\_1bit$ the number of faults for the 1-bit adder which equals 32. This result is used to evaluate the achieved fault coverage. The other columns report how the detection capability varies with respect to the different overloading strategies for the + operator. Each entry shows the achieved fault coverage, that is the number of times the methodology guarantees that the result is either correct or an error signal is raised. For example, using 2-bit operands, the adoption of the control operation op2=ris-op1 does not allow the detection of a fault causing an observable error in 32 situations, a fault coverage of 96.88%. The number of undetected errors drops, though implying a higher overhead in terms of computational time and/or area overhead.

Another interesting measure consists of the capability of the methodology to detect the fault, independently of the fact that it produces an erroneous result or not (i.e. independently of the fault observability). By analyzing the 2-bit adder case, the number of observable errors is 216 yet in almost all approaches the technique allows fault detection also when the produced result is correct: 352 (out of 1024) for "Tech1", 384 for "Tech2" and 428 for the combination of both. This property allows the reduction of the probability of having a second fault occur before the first one is detected, thus improving the system reliability. It is worth mentioning that classical Self-Checking design techniques provide fault detection capabilities only when the fault is observable. In conclusion, the example shows that for a ripple-carry adder (independently from its actual implementation) using different functional units to perform related operations the coverage is always 100%; using the same unit the percentage of input combinations that by-pass the checks varies, depending on the adopted strategy, in the range $[81.90\%, 99.87\%]$.

The same approach has been applied to the other operators (-, *, and /), analyzed w.r.t. several overloading techniques and their combination, achieving similar results in terms of fault observability and fault coverage.

## 5 The reliable co–design flow

This section shows the reliable co-design flow based on self-checking specification used to evaluate cost and performance of the proposed approach. Starting from a self-checking specification in *SystemC-Plus*, the design flow adopted (the flow is *OFFIS* [18] and *Synopsys* [16] based for the HW part, *GNU* [7] based for the SW one) is shown in Figure 3. By means of the *OFFIS SystemC-Plus Synthesizer*, *SystemC-Plus* is transformed into synthesizable behavioral and/or RT *SystemC*. By means of the $g$++ compiler we can easily obtain the software implementation, with *Synopsys CoCentric* the hardware one.

### 5.1 Methodology evaluation

By implementing both the reliable and not-reliable system specification a comparison in terms of costs and performance has been carried out, to evaluate the quality of the proposed approach on the final system. A FIR circuit has been designed with the proposed design flow [3], allowing an in-depth analysis of the approach. Several experiments have been carried out on the same circuit and results are reported in Table 3, while other circuits are now taken into consideration.

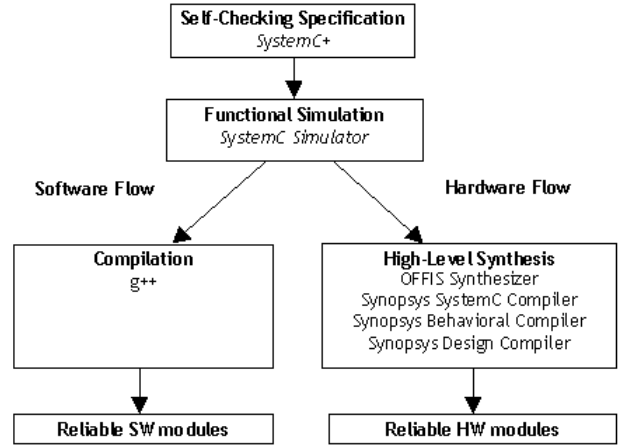

**Figure 3. The reliable co-design flow.**

|  |  | Hardware Implementation | |
|---|---|---|---|
|  |  | Latency (clock cycles) | CLB Slices |
| FIR | min area | $2 + 7n$ @ 20MHz | 412 |
|  | min latency | $2 + 5n$ @ 20MHz | 477 |
| FIR with SCK | min area | $2 + 10n$ @ 16.67MHz | 1926 |
|  | min latency | $2 + 5n$ @ 20MHz | 1593 |
| FIR embedded SCK | min area | $2 + 9n$ @ 15.38MHz | 634 |
|  | min latency | $2 + 5n$ @ 20MHz | 861 |

|  | Software Implementation | |
|---|---|---|
|  | Exe Time (sec) | Exe Size (KB) |
| FIR | 6.83 | 889 |
| FIR with SCK | 10.02 | 893 |
| FIR embedded SCK | 7.90 | 889 |

**Table 3. Application of the proposed methodology to the FIR.**

Results on this example are encouraging as far as the overloading mechanism is concerned, especially since such an approach has been used in the past but for different environments and purposes. The most significant aspects are:

- **extensible reliability library**: the overloading mechanism provides a flexible and extensible way to define a library or readily-available Self-Checking designs for the basic operators, each one with a cost/fault coverage characterization; the designer can select different self-checking approaches depending on the trade-off.
- **information hiding**: the most complex aspects related to reliability are transparent to the designer, for both hardware and software implementations.
- **standard technology**: the use of a mechanism part of the specification language allows the approach to be integrated in any SystemC hw/sw co-design flow.

On the other hand, the application of the approach pointed out the possibility of incurring in **template conflicts** when templates for overloading are already part of the original description. In such a case the proposed approach cannot be adopted.

This last motivation, together with the desire to be independent of proprietary software (such as the converter from SystemC-Plus to SystemC), is leading to the evolution of the proposed approach to virtually exploit the overloading mechanism directly in the SystemC description.

When the software implementation is selected, analyses have been carried out to verify that the redundant operations for achieving the desired reliability are not "simplified" by the compiler thus nullifying the operator overloading efforts. Both code size and execution times remain almost unmodified.

Let us remember, that the achieved fault coverage is complete for hardware implementation, whereas software implementations are characterized by the coverage estimated for the single functional unit, where the lowest estimated value is about 98%.

## 6 Concluding remarks

The paper proposes a methodology for designing circuits with reliability properties by introducing in the initial system specification elements able to provide the ability to autonomously detect the occurrence of hardware faults. The approach provides such on-line testing capabilities by using the overloading mechanism for the *SystemC-Plus* language, today's specification language for hardware/software co-design flows. In a manner that is completely transparent to the designer, "hidden" operations perform additional computation to verify the correctness of the resulting data. The achieved results in terms of fault coverage are promising and future work will focus on the trade-off between fault coverage and costs, in order to allow the designer to select the desired level of reliability while keeping area overhead and performance degradation within an acceptable limit.